\documentclass[showpacs,showkeys,twocolumn]{revtex4}
\usepackage{graphicx}
\usepackage{hyperref}
\usepackage{amssymb}
\usepackage{feynmp}

\def\beq{\begin{equation}}
\def\eeq{\end{equation}}

\begin{document}

\title{Enhanced Emission and Accumulation of Antiprotons and
Positrons from Supersymmetric Dark Matter}

\author{Chen Jacoby}
\email{chj3@post.tau.ac.il}
\affiliation{School of Physics and
Astronomy, Tel Aviv University Ramat-Aviv, Tel Aviv 69978, Israel}
\author{Shmuel Nussinov}
\email{nussinov@post.tau.ac.il}
\affiliation{School of Physics and
Astronomy, Tel Aviv University Ramat-Aviv, Tel Aviv 69978, Israel}

\begin{abstract}
We estimate the amount of antiprotons and positrons in cosmic rays
due to neutralino annihilations in the galactic halo assuming that
dark matter tends to cluster and that these clusters are not
disturbed by tidal forces. We find that, assuming neutralinos
annihilate mostly to gauge bosons, the amount of antiprotons
should exceed the number seen at BESS, whereas the increase in
positron flux is below the present detection threshold.
\end{abstract}



\preprint{hep-ph/0612065}

\preprint{TAUP-2842/06}

\maketitle

\section{Introduction} \label{sec:introduction}
The nature of dark matter remains one of nature's puzzles.
Supersymmetry with conserved $R$-parity, besides having many
desirable properties for particle physicists, provides a natural
candidate for cold dark matter, since if $R$-parity is conserved,
then the lightest supersymmetric partner is stable. In most
models, the lightest supersymmetric partner is the lightest
neutralino, a linear combination of the neutral supersymmetric
partners: the Bino $\tilde{B}$, the Wino $\tilde{W}^3$ and the two
neutral Higgsinos $\tilde{h}_1$ and $\tilde{h}_2$.

Detection of supersymmetric dark matter can be achieved indirectly
through its effect on cosmic rays, and more specifically on
antiprotons and positrons in the cosmic rays
\cite{Rudaz:1987ry,Bottino:1994xs,bottino:1998tw,Donato:2003xg,Bottino:2005xy,Cumberbatch:2006tq,Lavalle:2006vb,Bi:2006vk,deBoer:2006tv,Yuan:2006ju}.
(See also ref.
\cite{Zeldovich:1980st,Fargion:1994me,Fargion:1999ss,Belotsky:2004st}
for discussion on the effects of WIMPs on cosmic rays).

Recent $N$-body numerical simulations suggest that the first
structures that were formed in the universe at $z=26$ were dark
matter mini-halos of mass $10^{-6}M_\odot$ and half mass radius of
$0.01 \,$pc \cite{Diemand:2005vz}. If these mini-halos survive
disruptive gravitational forces in the galaxy, the enhanced dark
matter density inside the mini-halos enhance the neutralino
annihilation rate.

Our goal in this paper is to determine the possible effects such
an enhanced abundance has on the emission of antiprotons and
positrons and on the detection of these antiparticles in cosmic
rays.

In section \ref{sec:clustering} we give a short review on recent
results regarding the clustering of cold dark matter. Section
\ref{sec:susy} introduces the supersymmetric framework in which we
work. The production of antiprotons due to neutralino annihilation
is discussed in section \ref{sec:production} and the antiprotons'
propagation through the galaxy is presented in section
\ref{sec:propagation}.

Finally we discuss the implications of the enhanced neutralino
density on the detection of positrons in section
\ref{sec:positrons}.

\section{Clustering of Cold Dark Matter}
\label{sec:clustering}

Unlike hot dark matter with light constituents, which are
relativistic throughout most of the relevant history, or baryons,
which via electromagnetic interactions with the cosmic microwave
background radiation tend to maintain their temperature until late
times, cold dark matter clusters in the early universe, once the
perturbations $\delta(\rho)/\rho $ on the scale considered enter
the horizon. Structures then form in a down-up hierarchial
fashion:  smaller structures form first and merge into bigger
ones.

This culminates in forming galaxies with local dark matter density
in our neighborhood of \beq  \rho_\chi^{local} \sim  0.4 \frac
{GeV}{cm^3}, \
  \eeq
clusters thereof and even larger structures.

At some red-shift $z$, certain autonomous structures form in the
initially almost smooth cold dark matter \cite{Green:2005fa}.
Those become gravitationally bound and decouple (except for the
center of mass motion) from the Hubble expansion. The local dark
matter density within these first mini-halos, is enhanced relative
to the average co-moving number density, i.e, the present
cosmological dark matter density, and becomes \beq
n_\chi^{mini-halo}\sim  (6z)^3 \cdot
 n_\chi^{cosmological}. \eeq

The factor $z^3$ reflects the hubble stretching and $n_\chi$
dilution which operates cosmologically but not inside the
mini-halos and the $6^3$ factor is due to the extra factor $6$
shrinkage of the virilized mini-halo.

Thus, if the first structures form at $z=100-20$ we have
enhancement factors of \beq e_f \sim  (6z)^3 \sim  2 \cdot 10^8- 2
\cdot 10^6 \ \eeq over the co-moving average density of
neutralinos.

Recent detailed N body simulations with initial fluctuations
extrapolated down from WMAP to galaxies to the $10^{-15}$ less
massive dark matter mini-halos suggest indeed  that
\cite{Diemand:2005vz}:

\begin{enumerate}
\item These first structures formed early on at z=100-20 so that
the neutralino concentration therein is enhanced by $ e_f \sim
2\cdot 10^8- 2\cdot10^6$  over the cosmological: \beq
 n_\chi^{cosmological} m_\chi = \Omega_\chi \rho_{C}\sim 1.3 \
 keV/cm^3. \label{dens} \eeq
 Hence, the neutralino mass density inside the mini-halo is approximately $3-300 \ GeV/cm^3$,
 i.e, $5-750$ times larger than the average dark matter density in our local
 neighborhood.

\item More specifically the masses of the smallest mini-halos were found to
be \beq m_{mini-halo}\sim  10^{-6} m_\odot, \eeq and sizes \beq
r_{mini-halo}\sim 0.01pc \sim  3 \cdot 10^{16} cm. \eeq The
average density inside these structures is approximately $40
$~GeV/cm$^3$, i.e,
 100 times that of the average local dark matter density. The density profile of these mini-halos can be described
 by a power law \beq \rho(r)\varpropto r^{-\gamma}, \eeq with
 $\gamma$ in the range from 1.5 to 2.

\item Let $f_{mini-halo}$ be the fraction of all the cosmological dark
matter clustered inside these first mini-halos. It is important
for our present purpose that subsequent incorporation
 of mini-halos alongside some of the background initially unclustered dark matter, into later, bigger
 mini-halos, keeps the smaller, more compact, first mini-halos intact.
 The authors of ref. \cite{Diemand:2005vz} find that this is indeed the case.  More specifically, tidal
 disruption by stars and by the collective gravitational field does not destroy the
 above first mini-halos even inside the galactic disc at distances larger than $3 \,$kpc from
 the center of our galaxy and, in particular, at our location $7 \, $kpc from the
 center. The question of the stability of these structures has been further discussed in \cite{Zhao:2005py,Moore:2005uu,Berezinsky:2005py,Green:2006hh,Goerdt:2006hp}
\end{enumerate}

The actual value of $f_{mini-halo}$ is estimated to be
approximately $0.05$. In the galaxy and in the halo the mini-halos
track the conventional ordinary unclustered dark matter
distributions
 $ \rho (r) \varpropto  1/r^2$. The authors of \cite{Diemand:2005vz} estimate an average separation between
 the early compact mini-halos in our neighborhood of \beq d\sim0.1
\, pc, \eeq or more precisely, $500$ mini-halos within a $(pc)^3$.

 This is equivalent to a local average density of \beq
 500\cdot 10^{-6} m_{\odot} {{pc}^{-3}}\sim  \frac{1}{20} \cdot 0.4 \,GeV/cm^3,\eeq
namely, approximately $5\%$ of the local dark matter density,
which implies equal enrichment in the
 galaxy of the unclustered neutralinos and the initial mini-halos. This seems a conservative estimate
 of the density of the initial mini-halos as larger structures are likely to form near the
 previous, slightly smaller structures.

\section{The Supersymmetric Model}
\label{sec:susy}

We work here in the framework of $N=1$ minimal supergravity
\cite{Drees:1992am, Djouadi:2006be,Allanach:2005kz}.

For our purposes, it is sufficient to describe the model using
only four parameters and a sign. These parameters are $m_0$, the
common scalar mass, $m_{1/2}$, the common gaugino mass and $A_0$,
the common trilinear interaction term, all of which are evaluated
at the GUT scale. The fourth parameter is $tan\beta$, the ratio
between the vacuum expectation values of the two Higgs doublets,
and the final parameter is the sign of $\mu$, the parameter that
appears in the Higgs superfields interaction.

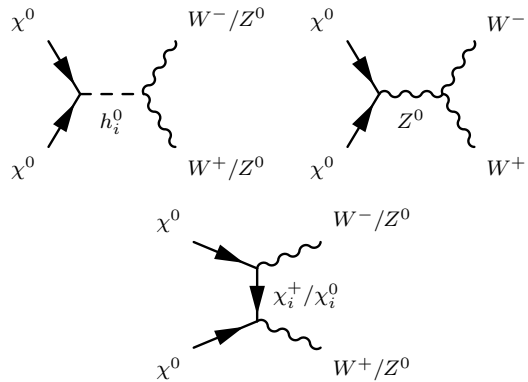
\begin{figure}[!htb]
\center{
\begin{fmffile}{annihilation}
\fmfframe(5,10)(25,5){
\begin{fmfgraph*}(60,40)
    \fmfleft{i1,i2}
    \fmfright{o1,o2}
    \fmflabel{\footnotesize{$\chi^0$}}{i1}\fmflabel{\footnotesize{$\chi^0$}}{i2}
    \fmflabel{\footnotesize{$W^+/Z^0$}}{o1}\fmflabel{\footnotesize{$W^-/Z^0$}}{o2}
    \fmf{fermion}{i1,v1}
    \fmf{fermion}{i2,v1}
    \fmf{dashes,label=\footnotesize{$h^0_i$}}{v1,v2}
    \fmf{boson}{v2,o1}
    \fmf{boson}{v2,o2}
\end{fmfgraph*}}
\fmfframe(25,10)(5,5){
\begin{fmfgraph*}(60,40)
    \fmfleft{i3,i4}
    \fmfright{o3,o4}
    \fmflabel{\footnotesize{$\chi^0$}}{i3}\fmflabel{\footnotesize{$\chi^0$}}{i4}
    \fmflabel{\footnotesize{$W^+$}}{o3}\fmflabel{\footnotesize{$W^-$}}{o4}
    \fmf{fermion}{i3,v3}
    \fmf{fermion}{i4,v3}
    \fmf{boson,label=\footnotesize{$Z^0$}}{v3,v4}
    \fmf{boson}{v4,o3}
    \fmf{boson}{v4,o4}
\end{fmfgraph*}
} \fmfframe(5,30)(5,5){
\begin{fmfgraph*}(60,40)
    \fmfleft{i5,i6}
    \fmfright{o5,o6}
    \fmflabel{\footnotesize{$\chi^0$}}{i5}\fmflabel{\footnotesize{$\chi^0$}}{i6}
    \fmflabel{\footnotesize{$W^+/Z^0$}}{o5}\fmflabel{\footnotesize{$W^-/Z^0$}}{o6}
    \fmf{fermion}{i5,v5}
    \fmf{fermion}{i6,v6}
    \fmf{fermion,label=\footnotesize{$\chi^+_i/\chi_i^0$}}{v6,v5}
    \fmf{boson}{v5,o5}
    \fmf{boson}{v6,o6}
\end{fmfgraph*}
}
\end{fmffile}}
\caption{Neutralino annihilation into two gauge bosons}
\label{annihilation}
\end{figure}

We assume that the mass of the lightest neutralino is larger than
the mass of the $W$ and the $Z$ bosons, but either smaller or not
much larger than the mass of the lightest Higgs particle. If this
is the case, and assuming that the neutralino is mostly Wino-like
or Higgsino-like, annihilation of neutralinos will be dominated by
annihilation to two gauge bosons (fig. \ref{annihilation}).

The authors of ref. \cite{Ellis:2005tu} estimate that the mass of
the lightest neutral higgs boson  most likely lies within the
range \beq 114 \ GeV < m_{h_1} < 127 \ GeV, \eeq and thus the mass
of the lightest neutralino will be taken as \beq m_\chi=\eta\cdot
100 \ GeV, \label{eta}\eeq with $\eta \gtrsim 1$.

The next to lightest supersymmetric particle is assumed to be much
heavier (several hundreds of GeV), and thus effects of
co-annihilations involving almost degenerate supersymmetric
particles can be neglected.

\section{Production of Antiprotons from Neutralino Annihilation}
\label{sec:production}

  Our discussion of the enhanced flux of antiparticles is based on the
 standard mini-halos of mass  $m=10^{-6} m_{\odot}$, and size $r \sim   0.01$ pc \cite{Diemand:2005vz}.

  We assume a local mini-halo density in our galactic neighborhood
  of $500 \ $pc$^{-3}$. The average neutralino mass density in such a mini-halo is
\beq
   \rho_{\chi} \sim 10^{57} \frac {GeV}{pc^{3}},
\eeq and a number density of

\beq n_{\chi} \sim \eta^{-1} \cdot 10^{55} {pc^{-3}}, \eeq where
$\eta$ is defined in eq. (\ref{eta})

 The lifetime for $\chi\chi$ annihilations within these mini-halos is
\beq \tau_{ann}=\left(\left<\sigma_{ann} v \right>n_\chi
\right)^{-1} \sim 3 \cdot 10^{26}\eta \ sec, \eeq where \beq
\left<\sigma_{ann} v \right>=10^{-9}GeV^{-2}. \eeq

The differential production rate per unit of volume of antiprotons
can be written as

\beq q_{\bar{p}}=\left<\sigma_{ann} v\right>
g\left(T_{\bar{p}}\right)\left(\frac{\rho_\chi}{m_\chi}\right)^2,
\eeq where $g\left(T_{\bar{p}}\right)$ denotes the antiproton
differential energy spectrum.

Assuming that neutralinos annihilate mostly to gauge bosons, one
antiproton is produced on average per annihilating neutralino
 through decay of $W$ or $Z$ bosons \cite{Eidelman:2004wy}, leading
to about $10^{-27}$ antiprotons per cm$^3$ per second which are
injected in the galactic disc and in the halo in our neighborhood.

  Let the effective residence time of the antiprotons before slowing down and
 annihilating be \beq t_{res}= \tau_{r} \cdot 10^7 years\sim  \tau_r \cdot
3\cdot10^{14} sec. \eeq
  The antiprotons' density builds up to
\beq
 n_{\bar{p}}\sim  \eta^{-1} \cdot \tau_r \cdot 3 \cdot 10^{-13} cm^{-3} \eeq
  The antiprotons of interest with energy above $1 \ GeV$ move with velocities $v \approx c$ making a flux:
  \beq
  \Phi_{\bar{p}} \sim  c\cdot n \sim   \eta^{-1}\tau_r \cdot 10^{-2} cm^{-2}sec^{-1},\eeq
  yielding an angular differential flux of
  \beq \eta^{-1}\tau_r 8\cdot 10^{-4}\ {cm^{-2}
  sec^{-1}sr^{-1}}.\eeq

\begin{figure}[!htb]
\begin{center}
\includegraphics[width=8.6cm, height=8.6cm]{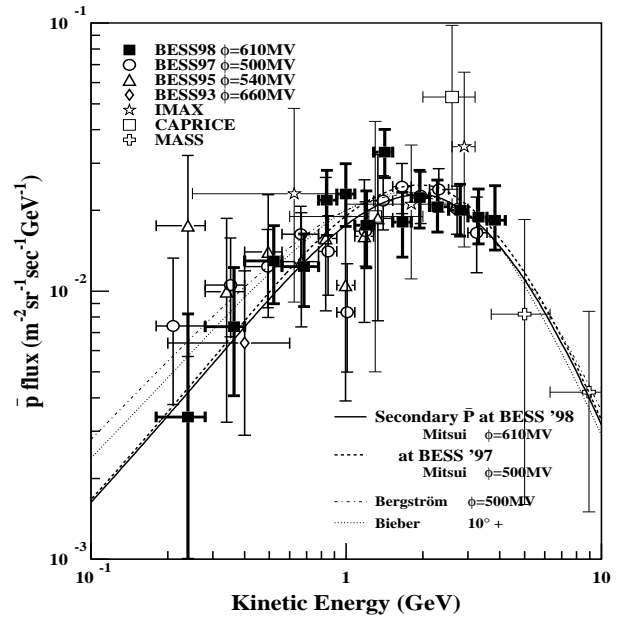}
\caption{Antiproton spectrum from Bess.} \label{fig:bess}
\end{center}
\end{figure}

   If, as we argue in section \ref{sec:propagation}, the residence time is indeed $10^7$ years, then the expected  range of
antiproton fluxes from neutralino annihilations in mini-halos
exceeds the experimental values seen at BESS
\cite{Asaoka:2001fv}(fig. \ref{fig:bess}).

The angular differential flux integrated from $0.1$ GeV to $10$
GeV using the data from BESS is \beq \sim 10^{-5} \,
cm^{-2}sec^{-1}sr^{-1} \eeq

  It has been known for some time that cosmic ray antiprotons are a sensitive indicator
 for supersymmetric dark matter. This is even more so  with the enhanced annihilation rates inside the
 mini-haloes in our galaxy. The local unclustered dark matter of $0.4 \ GeV/{cm^3}$
 in our neighborhood is $0.25 \cdot 10 ^6$ times
 that of  the cosmological $1.5 \ keV/{cm^3}$
 as opposed to about a $100$ times  larger enhancements inside the mini-halos.

 Since the fraction of neutralinos in our galactic neighborhood residing in mini-halos is $\sim  0.05$, the net
 effect of the local mini-halos is to enhance the antiproton flux (relative to the expectation for a
 uniform neutralino density of $0.4 \, GeV/cm^3$) by more than $100$. Thus, independently of any detailed estimate
 of the actual antiproton flux in the two scenarios, the neutralino clustering within mini-halos can help
 the signal of antiprotons from neutralino annihilations cross the detectibility threshold.

\section{Propagation of Antiprotons} \label{sec:propagation}

  Let us briefly review the estimate of the lifetime of the antiprotons. The antiprotons, like other
 components of the cosmic rays, can disappear from the reservoir of cosmic rays in two ways:

\begin{enumerate}
\item  by literally "leaking" out into the intergalactic space along
magnetic field lines  which do not close inside the galaxy or the
halo.

\item While traversing the disc and to some extent also in the halo,
the antiprotons lose energy via collisions with electrons and
protons in the ambient plasma. Such antiprotons with too low an
energy cannot penetrate the stelar wind and attendant magnetic
fields and thus do not arrive at (ant)arctic balloons with minimal
geomagnetic cutoff.
\end{enumerate}
 In the special case of  antiprotons, losses via annihilation are most important. Indeed using:
\beq \sigma_{\bar{p}-p}^{ann}v \sim 10^{-25} cm^2, \eeq we find
the the lifetime for annihilation is
 \beq
\tau^{ann}_{\bar{p}}= (n_{p} \cdot \sigma_{\bar{p}-p}^{ann}v)^{-1}
\sim 10^7\,years, \eeq and thus, the total traversed grammage is
\begin{eqnarray}
\sim 10^{25} protons/cm^2 \sim 10 gr/cm^2.
\end{eqnarray}

The magnetic fields in the disc which are of the order of $3 \cdot
10^{-6}$\,gauss are stronger by a factor of about $10$ than those
in the halo, and the latter are a $100$ times larger than the
intergalactic magnetic fields. Thus only a small fraction of the
magnetic field lines in the disc connect to the halo and  a
smaller fraction yet of the magnetic fields in the halo connect to
the outside thereby slowing down the leakage. Indeed the overall
residence time (in the disc and the halo together) of protons with
energy of the order of $1\,$GeV is estimated using isotopic
composition data to be $30$ millions years, allowing $1000$
traversals of the disc or $100$ traversals of the halo.

We also need to restrict the total grammage traversed by the
cosmic ray particles to be less than $5 gr/cm^2$. This yields:
\beq t_{disc}\sim  5 \cdot 10^6 \textrm{years},\eeq namely, \beq
\tau_r\sim 1/2,\eeq roughly consistent with $10^7$ years (or
$\tau_r \sim 1$) used in estimating the antiproton signal from
neutralino annihilation.

 Note that in the present scenario antiprotons are produced not only in the
 disc  as most cosmic rays but even more within the mini-halos. This enhances the signal
 as antiprotons originating in the halo can go into the disc and arrive to us as
 well.

  The following comments are in order:

\begin{enumerate}
 \item The estimated antiproton flux is reduced by a factor of about $0.7$, the fraction of hadronic $Z^0$
 (and/or
 $W^+$) decays.  With the smaller \mbox{$\tau_r\sim 1/2$} this reduces the expected flux by
 factor 3, yielding a differential antiproton flux of \beq 2.7
 \cdot 10^{-4} \ cm^{-2}sec^{-1}sr^{-1}.\eeq
 \item  By and large, the observed antiproton signal agrees with that computed from interaction of
 cosmic ray protons with momenta $\gtrsim  8$ Gev with interplanetary protons suggesting more
 stringent limits on the supersymmetric dark matter scenario from antiprotons.  The background
 antiprotons are relatively energetic. It was therefore suggested that the excess at
 energies lower than a GeV indicates supersymmetric dark matter annihilations.
\item As noted above, the multiplicity of antiprotons produced (particularly those at the more
interesting lower energies) grows rapidly with the mass of the
neutralino if the mix of Wino, Bino and Higgsino in $\chi^0$
optimizes the annihilation via a virtual   $Z^0$ (and to $Z^0
\rightarrow \bar {q}q$, in particular). This will enhance the
desired signal of antiprotons from neutralino annihilations. On
the other hand, if annihilation to $W^+ W^-$ dominates, the
multiplicity is fixed but the energy spectrum shifts upward with
$m_\chi$, diminishing the signal of low energy antiprotons.
\item If the mass of the neutralino is large enough, annihilations
into either a higgs boson and a gauge boson or into two higgs
bosons are possible. The higgs bosons will then decay to $b$
quarks, and the production of antiprotons is suppressed.

\item The cross section for annihilation of antiprotons with
energies of the order of $1\,$GeV is larger than the corresponding
elastic cross section. Further, already at these energies the
momentum transferred in these scattering, $t$, and the attendant
recoil kinetic energy loss, $t/2m_{p}$, are relatively small. Thus
the detected antiprotons manifest the energy spectrum from
neutralino annihilations.
\end{enumerate}

\section{Neutralino Detection through Positrons} \label{sec:positrons}

We next consider positrons. Could their observation be a better
indicator for supersymmetric dark matter?

The answer is clearly negative in so far as the lower energy part
of the spectrum with energy less than $1\,$GeV is concerned. The
background from $pp \rightarrow pn \pi^+$ with the charged pion
decaying to muon and then to a positron is huge compared to the
antiproton background: first the threshold in this case is lower
and it is possible to utilize the lower and much stronger part of
the cosmic radiation spectrum. Also the inclusive pion production
cross section is about $1000$ times larger than that for
antiprotons. The multiplicity of positrons from neutralino
annihilation is larger by a factor of $15$ than that of the
antiprotons but this cannot compensate the above larger factors.

The situation is clearly better if we focus on the more energetic
positrons, for instance, those with $E_{e^+}\sim8\,$ GeV or higher
originating from charged pions of $E_{\pi}>20\,$ GeV.  The
multiplicity of those in neutralino annihilations is still
$\mathcal{O}(1)$ whereas their production in proton-proton
collisions requires now protons of energies greater than
$30\,$GeV. This is strongly suppressed by the power fall-off of
the cosmic rays spectrum. (The authors of ref.
\cite{Hooper:2004bq} tried to relate the apparent enhancement in
this part to a particular SUSY model. Here we focus more on the
overall magnitude of the signal).

An extra advantage is that the main factor shortening the
accumulation time of antiprotons in the galaxy, namely, their
large annihilation cross section, has no counterpart here: at
these high energies $e^+ e^-$ annihilations are negligible.

The various losses on some existing electromagnetic background,
proportional to $\gamma^2$, are $2.5 \cdot 10^8$ faster for the
energetic positrons than for the $1\,$GeV antiprotons. The actual
energy loss rate is \beq dW/{dt} = \gamma^2 \cdot
\sigma_{Thompson} \cdot U \cdot c \sim  5 \cdot 10^{-6} U,\ \eeq
where $U$ is the energy density due to the magnetic fields, and
\beq U = B^2/(8\pi) \sim 0.22 \, eV \cdot (cm)^{-3}. \eeq

Thus $8$\,Gev positrons lose $1/2$ of their energy via synchrotron
radiation in $3.5 \cdot 10^{15}$\,sec. This is more than ten times
longer than the lifetimes of the antiprotons. Roughly the same
loss occurs on the equal energy density cosmic microwave radiation
via inverse Compton scattering. Finally the standard Coulombic
losses are very similar for the minimally ionizing relativistic
positrons and antiprotons and in any case are less than those of
nuclear interactions and annihilations at these energies. Using a
similar analysis as in section \ref{sec:production} we get a
positron angular flux of \beq 0.4\,cm^{-2}sec^{-1}sr^{-1}.\eeq
This is an order of magnitude smaller than the observed amount.

The authors of ref. \cite{Cumberbatch:2006tq} assert that the
positron excess observed on both flights of the HEAT balloon
experiment \cite{Barwick:1995gv,Barwick:1997ig} can be explained
by a single nearby mini-halo. This scenario is very unlikely, as
shown in \cite{Lavalle:2006vb}.

\section{Conclusions}
In this paper we have shown that if neutralinos annihilate mostly
to gauge bosons there should be an excess of antiprotons over the
observed amount in BESS due to the enhanced neutralino density in
the mini-halos over the cosmological density. This fact may imply
that the mass of the neutralinos is larger than the mass of the
lightest higgs boson, in which case annihilations into higgs
bosons suppress the number of antiprotons produced, or that it is
smaller than the mass of the gauge bosons.

We have also shown, that although antiprotons are strong
indicators for dark matter, this is not the case for positrons.
The low energy positrons coming from neutralino annihilation
cannot be distinguished from the large background signal and the
contribution of neutralino annihilation to the positron signal
with high energies is too small to make any predictions.

\begin{acknowledgements}
We would like to thank Rennan Barkana for discussions that helped
initiate this project, to Yoel Rephaeli for discussions of other
possible manifestations of the mini-halos and also to Joseph Silk
for further discussions of the positron signature.
\end{acknowledgements}

\bibliographystyle{apsrev}
\bibliography{C:/temp/lsp}

\end{document}